\documentclass[aps,prb,twocolumn,longbibliography,10pt,superscriptaddress,
]{revtex4-1}

\usepackage{amsmath,amssymb,graphicx}
\usepackage{braket}
\usepackage{multirow}
\usepackage{xfrac}
\usepackage{bm}
\usepackage{url}
\usepackage{hyperref}
\hypersetup{
	colorlinks=true,
	linkcolor=blue,
	citecolor=blue,
	filecolor=magenta,
	urlcolor=cyan,
	breaklinks=true
}

\begin{document}

\title{Low-dimensional quantum gases in curved geometries}

\date{\today 
}

\author{Andrea Tononi}
\email{andrea.tononi@universite-paris-saclay.fr}
\affiliation{Universit\'e Paris-Saclay, CNRS, LPTMS, 91405 Orsay, France}
\author{Luca Salasnich}
\affiliation{Dipartimento di Fisica e Astronomia ’Galileo Galilei’, Universit{\`{a}} di Padova, via Marzolo 8, 35131 Padova, Italy}
\affiliation{Padua Quantum Technologies Research Center, Universit{\`{a}} di Padova, via Gradenigo 6/b, 35131 Padova, Italy}
\affiliation{INFN - Sezione di Padova, via Marzolo 8, 35131 Padova, Italy and CNR-INO, via Carrara 1, 50019 Sesto Fiorentino, Italy}

\begin{abstract}
\noindent 
Atomic gases confined in curved geometries {are characterized by} distinctive features that are absent in their flat counterparts, such as periodic boundaries, local curvature, and nontrivial topologies. 
The recent experiments with shell-shaped quantum gases and the study of ring-shaped superfluids point out that the manifold of a quantum gas could soon become a controllable feature, thus allowing to address the fundamental study of curved many-body quantum systems. 
Here, we review the main geometries realized in the experiments, analyzing the theoretical and experimental status on their phase transitions and on the superfluid dynamics. 
In perspective, we delineate the study of vortices, the few-body physics, and the search for analog models in various curved geometries as the most promising research areas. 
\end{abstract}

\maketitle

\section{Introduction}
\noindent
Investigating quantum many-body physics in curved spatial geometries is a new research trend in the field of ultracold atoms, driven by the recent experimental realization of shell-shaped gases \cite{carollo2022,jia2022}, and by a long tradition of studies on ring-shaped superfluids \cite{amico2022}. 
While the models of infinite systems were usually adopted to describe gases confined in flat boxes \cite{navon2021}, or the local physics of three-dimensional trapped configurations\cite{dalfovo1999}, it is now emerging the idea to include the system geometry directly in the theoretical description. 
The main goal of the research on curved atomic gases is taking advantage of the spatial configuration of the atoms to engineer new equilibrium and nonequilibrium phenomena that are inherently characteristic of the curved setting. 

In the last years, many papers have focused specifically on shell-shaped atomic gases, analyzing with different aims and techniques the phenomena of Bose-Einstein condensation and of superfluidity \cite{tononi2019,tononi2022}, and the collective excitations of the shells \cite{lannert2007,sun2018,padavic2018}. 
This theoretical attention is the product of a lively experimental scene aimed at the production of atomic bubbles \cite{carollo2022,jia2022} and of quantum gases confined on curved two-dimensional manifolds \cite{guo2022,rey2022}. 
More in general, the unique dynamics of vortices on superfluid surfaces with peculiar topologies has received a great deal of attention \cite{guenther2017,guenther2020,caracanhas2022}, and point out the necessity of finding a unified framework for describing vortex motion in a generic two-dimensional setting. 
Concerning ring-shaped gases, even if a one-dimensional confinement has not been experimentally demonstrated so far, various aspects pertaining to the curved geometry were recently analyzed, for instance investigating the analogy between the phononic excitations in expanding rings and cosmological phenomena\cite{eckel2018}, or the formation of a dynamical ring on a rapidly-rotating curved surface \cite{guo2020}. 

This Perspective aims to consolidate the field of quantum gases in curved geometries by reviewing the fundamental results and by suggesting new research directions. 
We begin by contextualizing the concepts of curvature and geometry for a system of ultracold atoms, identifying the innovative features with respect to the flat configurations. 
Then, we discuss the main experimental solutions adopted to confine the atoms on shell-shaped and on ring-shaped manifolds, and we briefly review the main theoretical results obtained in these geometries. 
In the final section, we conclude by outlining the prospects for future theoretical research, framing the analysis of various curved geometries along a few directions: the dynamics of vortices, the few-body physics, and the search for analog models with curved quantum gases.
\\

\begin{figure*}[hbtp]
\resizebox{0.618 \textwidth}{!}{\includegraphics*{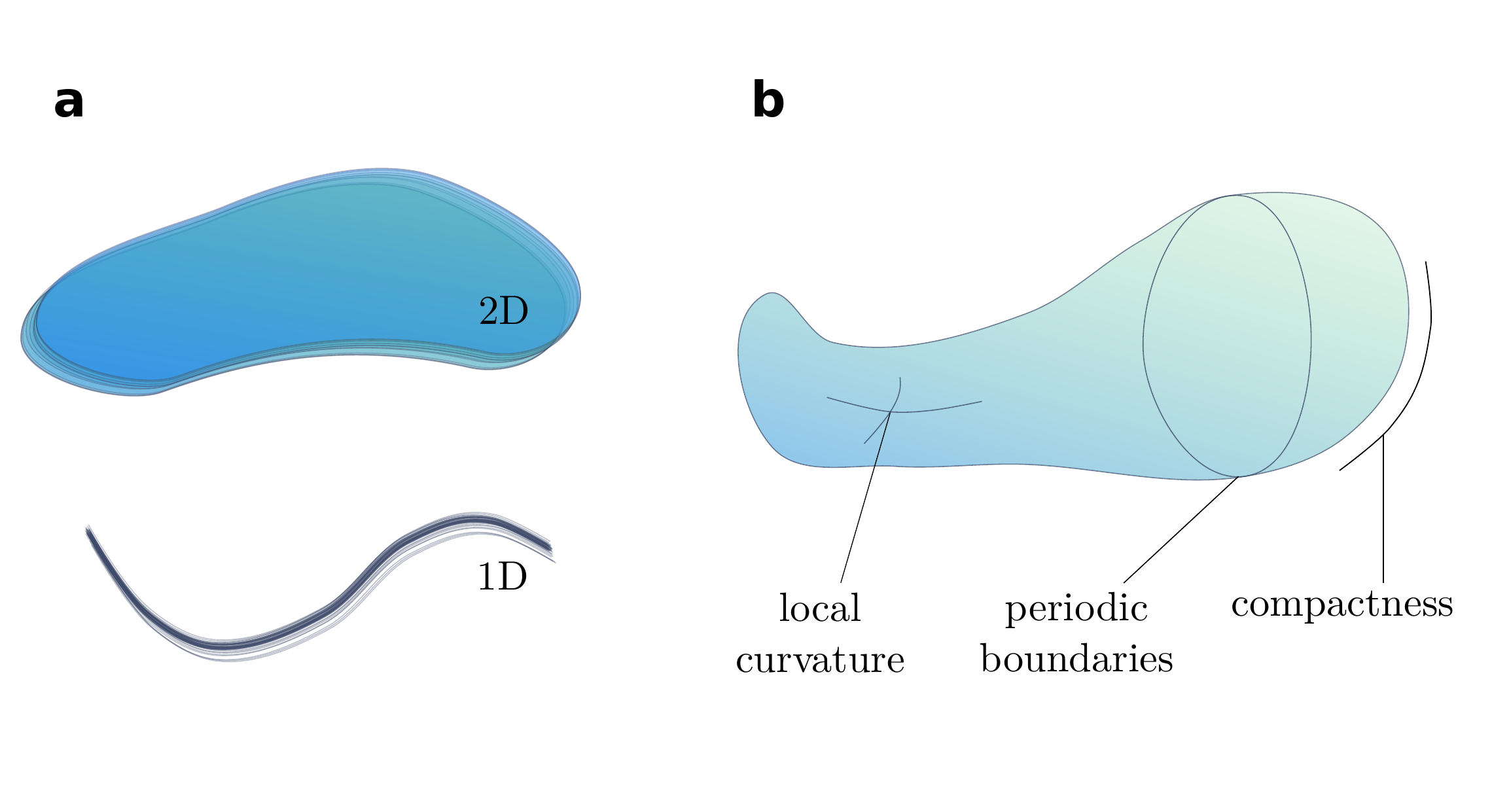}}
\caption{{\bf Curved geometries.} \textbf{a} \textbar \ A low-dimensional spatial configuration is obtained by confining the elements of a system near a 2D or a 1D manifold. 
When the local direction of the external confinement varies in space, the atoms of the system are effectively moving in a curved geometry. 
\textbf{b} \textbar \ 
Quantum gases in curved geometries may display distinctive features with respect to their flat counterparts, such as: periodic boundaries, a locally varying curvature, and compactness of the manifold corresponding to nontrivial topologies, such as the sphere, or the torus. 
} \label{fig1}
\end{figure*}

\section{Curvature and geometry}
\noindent
Quantum gases can be confined in geometries such as rings, shells and cylinders by designing appropriate potentials using external magnetic and electric fields. 
The motion of the atoms is effectively taking place on a low-dimensional manifold when, with respect to the characteristic energy of the transverse confinement, the thermal and the interaction energies of the gas are sufficiently weak\cite{pitaevskii2016}.
Under this condition, two mutually-independent features 
of the curved geometry can be identified {(FIG.~\ref{fig1}b)}, which distinguish curved quantum gases from their flat counterparts. 
The first is that boundary conditions for the atomic motion are generally nontrivial (periodic, for instance), in contrast with those of flat gases and of three-dimensional configurations. The second is that the atoms can be constrained to move on a manifold with locally-varying curvature. 

Regarding the first aspect, while 
the study of systems with peculiar boundary conditions is widespread in physics, these are often tools for theoretical computations rather than explicit models of curved geometries. 
While the choice of boundaries is usually irrelevant for systems at the thermodynamic limit, atomic gases are, under this aspect, finite quantum systems. 
For instance, the spatial distribution of the flow of a two-dimensional superfluid depends on the manifold periodicity, and may differ substantially in geometric surfaces with the topology of the sphere\cite{caracanhas2022} or of the torus\cite{guenther2020}. 

Secondly, the local curvature of the manifold influences the longitudinal physics, even for very strongly confined gases, due to the appearance of a geometric potential in the low-dimensional effective description of the system\cite{dewitt1957,dacosta1982}.
In two dimensions (2D), the geometric potential can be written in terms of the two principal curvatures $\kappa_1$ and $\kappa_2$ of the surface\cite{jost2010} or, equivalently, in terms of the Gaussian curvature $K = \kappa_1 \kappa_2$ and the average curvature $H=(\kappa_1+\kappa_2)/2$. 
In one dimension (1D), similarly, it depends on the local curvature of the waveguide\cite{leboeuf2001,cozzini2006,sandin2017,salasnich2022}, and it is thus proportional to the inverse squared of the local curvature radius. 

The consequences of the geometric potential have not been observed yet since the energy scale associated to the manifold curvature is, in typical experimental situations, subleading with respect to trap inhomogeneities and to the chemical potential.
When the geometric potential is neglected in the theoretical description, one is basically solving a problem in reduced coordinates, without considering the procedure of dimensional reduction and ignoring the embedding of the gas into {the} three dimensional space. 
The only property characterising the physics of the gas in a curved geometry is, in this case, the presence of periodic boundaries and their interplay with the microscopic physical processes.
\\

\begin{figure*}[hbtp]
\resizebox{0.99 \textwidth}{!}{\includegraphics*{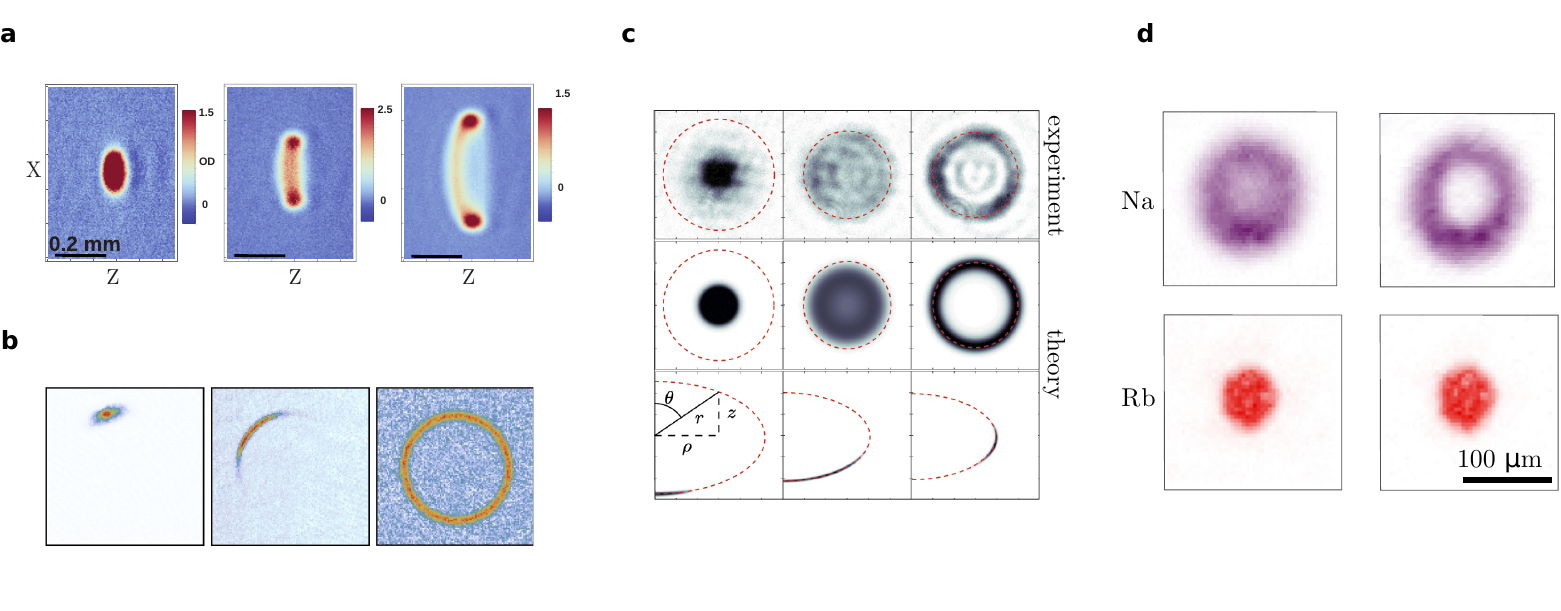}}
\caption{{\bf Experiments with {atomic} gases in curved geometries.}
\textbf{a} \textbar \ Ultracold gases have been confined in shell-shaped quasi two-dimensional configurations in recent experiments\cite{carollo2022} that implemented radiofrequency-induced adiabatic potentials in microgravity conditions. 
Harmonically-trapped condensates were inflated into $\sim 100 \, \mu\text{m}$-sized ultracold atomic bubbles by adiabatically changing the radiofrequency.
\textbf{b} \textbar \ By accelerating (first panel) and releasing (second and third panels) a Bose-Einstein condensate in a smooth magnetic waveguide, the regime of coherent hypersonic propagation in the curved ring-shaped geometry was achieved and maintained over a $15 \, \text{cm}$ distance\cite{pandey2019}.
\textbf{c} \textbar \ Experiments\cite{guo2022} with radiofrequency-induced adiabatic potentials exploited a inhomogeneous radiofrequency coupling to counterbalance the gravitational potential, thus displacing the atoms from the bottom of a shell (first column) to a curved-bowl configuration (second column) and then to a curved ellipsoidal band (third column). 
\textbf{d} \textbar \ The first observation of a shell-shaped condensate was recently achieved\cite{jia2022} by optically trapping with the same harmonic frequency a repulsive dual-species mixture of $^{87}$Rb, which occupies the central core, and of $^{23}$Na, which constitutes the thin shell around the core. The figures of the two columns are obtained with different imaging techniques. 
Part \textbf{a} adapted with permission from ref.\cite{carollo2022}, Springer Nature. 
Part \textbf{b} adapted with permission from ref.\cite{pandey2019}, Springer Nature. 
Part \textbf{c} adapted from ref.\cite{guo2022}, IOP. 
Part \textbf{d}
adapted with permission from ref.\cite{jia2022}, APS.
\\
} \label{fig2}
\end{figure*}

\section{Experimental status}
\noindent
The experiments with quantum gases of the last two decades have realized various curved geometries (FIG.~\ref{fig2}). 
We review the most significant implementations, focusing mainly on magnetic potentials for producing rings and shell-shaped superfluids, and on the discussion of repulsive atomic mixtures. 
In reviewing systems with real-space periodic boundaries, we set aside the complementary research on synthetic dimensions\cite{boada2012, boada2015, ozawa2019}, in which a particle degree of freedom can be used to mimic a periodic spatial coordinate, and where {the} local curvature is intrinsically absent. 

\subsection{Magnetic potentials for 2D geometries}
\noindent 
Atomic gases can be spatially confined by using proper configurations of magnetic fields. 
This can be understood already at a classical level: a static magnetic field $\mathbf{B}(\mathbf{r})$ can trap an atom with magnetic dipole moment $\bm{\mu}$ in the region of space near the minima of the potential energy $U(\mathbf{r})=-\bm{\mu}\cdot \mathbf{B}(\mathbf{r})$. 
This simple concept is the basis of harmonic {magnetic} potentials, whose strength grows quadratically with the distance from the trap center, but magnetic potentials can also be used to confine quantum gases in curved geometries. 
As shown in the early 2000s, ellipsoidal (shell-shaped) surfaces can be produced with the radiofrequency-induced adiabatic potentials\cite{zobay2000,zobay2001,zobay2004}.
The fundamental idea is to use a time-oscillating field in the radiofrequency range $\mathbf{B}_{\text{rf}}(\mathbf{r},t)$ to spatially modify (or ``dress'') the levels of a static magnetic potential: their proper combination leads the atoms to follow adiabatically a trajectory near the surface of an ellipsoid. 

Various experiments with these radiofrequency-induced adiabatic potentials have confined quantum gases in partially-filled (not closed) shells\cite{colombe2004, white2006,merloti2013,harte2018}, since the atoms accumulate to the lower side of the trap due to the gravitational potential. 
The first observation of ultracold atomic bubbles has been reported recently \cite{carollo2022}, and the key aspect which allowed to limit the detrimental role of gravity was performing the experiment in the Cold Atom Laboratory\cite{elliott2018,aveline2020,lundblad2019} (CAL), a space-orbiting facility on the International Space Station. 
The use of magnetic potentials is however not limited to {these space-based experiments}, and parallel efforts in Earth-based laboratories led to remarkable studies of quantum gases in curved geometries.
Recent works have analyzed rotating superfluids confined on a curved surface \cite{guo2020}, have produced curved superfluid bands by partially compensating the gravitational potential\cite{guo2022} (FIG.~\ref{fig2}c), and have considered the dressing of magnetic levels with multiple radiofrequency fields\cite{harte2018,luksch2019}.

In the future, the magnetic potentials could be used to engineer new geometries such as the surface of a torus\cite{fernholz2007}, a 2D manifold whose principal curvatures $\kappa_1$ and $\kappa_2$ can have opposite sign. 
Moreover, also cylindrical-shaped gases could be created with careful implementations of time-averaged adiabatic potentials\cite{lesanovsky2007,sherlock2011}, which consist in adiabatic potentials in which the static component of the magnetic field is properly averaged in time.
\\

\subsection{Superfluid rings} 
\noindent 
The study of quantum gases in ring-shaped geometries is vast and multifaceted, since this geometry is particularly suitable to study persistent superfluid flows, and can realize matter-wave circuits which have promising technological applications \cite{amico2021,amico2022}. 
Detailed reviews\cite{garraway2016} have already analyzed the implementation of ring-shaped gases using magnetic potentials, and we briefly highlight the most relevant research from the curved-geometry perspective.

In synthesis, after the first realizations of Bose-Einstein condensates (BEC) in magnetic waveguides \cite{gupta2005} {and the first observation of a persistent current\cite{ryu2007},} fully smooth ring-shaped traps based on time-averaged adiabatic potentials were first proposed theoretically \cite{lesanovsky2007} and later implemented experimentally \cite{sherlock2011}. 
Various experiments with optical potentials have addressed the superflow dynamics across a repulsive barrier located along the circuit \cite{ramanathan2011} and investigated the current-phase relationship with interferometric measurements \cite{eckel2014,eckel2014b}. 
Remarkably, atomic condensates were launched into painted optical potentials with dynamically-controllable shapes (waveguide bends, circuits, and Y-shaped junctions), and the propagation of the wave packet was then analyzed\cite{ryu2015}. 
Further relevant studies of the last few years involve achieving and exceeding the supersonic regime of superfluid propagation in a ring\cite{pandey2019} (FIG.~\ref{fig2}b), and creating persistent currents both in bosonic superfluids \cite{degoerdeherve2021} and in fermionic ones \cite{cai2022}. 

The experiments with rings conducted so far were implementing mostly 3D or 2D confinements, with the disadvantage that the longitudinal dynamics can be influenced by the asymmetry between the inner and outer edges\cite{moulder2012,dubessy2012}. 
Probing, for instance, the consequences of the non-constant local curvature along the ring, would first require the realization of a strictly one-dimensional ring confinement. 
By achieving this goal and by realizing schemes to produce waveguides and rings with locally-varying curvature\cite{ryu2015,sinucoleon2014}, it would then be possible to address experimentally long-standing analyses on the interplay between the curvature and the superfluid flow \cite{leboeuf2001}.
\\

\subsection{Repulsive atomic mixtures}
\noindent 
Also two-component quantum mixtures, obtained by confining together different atomic species, different isotopes, or different hyperfine states of the same atom, allow to study quantum gases in curved geometries. 
The basic idea is that two gases confined in a harmonic trap can separate in space, forming a inner core of one species surrounded by a shell of the other species \cite{ho1996,pu1998}. 
There are two crucial aspects for observing the shell configuration: the first is being in the regime of phase separation, i.e. the interspecies repulsion $g_{12}$ must be stronger than the geometric mean of the intraspecies repulsions $\sqrt{g_{11}g_{22}}$. 
Secondly, the gravitational potential yielding the acceleration of gravity $g$ must be counterbalanced\cite{wolf2022}, since it displaces the trap center\cite{derrico2019} of the two species by the different amounts $-g/\omega_1^2$ and $-g/\omega_2^2$, with $\omega_1$ and $\omega_2$ the harmonic confinement frequencies, and thus prevents the formation of a closed shell even for large atom numbers\cite{ospelkaus2006}.

There are currently two strategies to counterbalance gravity in repulsive atomic mixtures. 
The first possibility is to trap the atoms optically using a specific ``magic'' wavelength \cite{safronova2006,ospelkaus2006,onofrio2002}, which allows to trap the two species with the same harmonic frequency and with nearly overlapping center of mass\cite{jia2022}. 
This method has recently enabled the first observation of spherically-symmetric closed shells \cite{jia2022} (FIG.~\ref{fig2}d) and to study the self-interference of the atoms following the free expansion.
The second obvious possibility is to conduct the experiment in microgravity facilities, such as CAL\cite{elliott2018,aveline2020} and its further developments \cite{frye2021,thompson2023}, or {in} other free-falling setups \cite{vanzoest2010,condon2019}. 

There are many active experiments on two-component mixtures \cite{wille2008, voigt2009, hara2011, ferrierbarbut2014, wang2016, roy2017, cabrera2018, burchianti2018, semeghini2018, ravensbergen2018, derrico2019, neri2020, green2020}
in which shells, cylinders and other shapes could be produced by implementing these or other gravity-compensation mechanisms. 
The main challenges ahead will be producing quasi-two-dimensional shells with a sufficiently high atomic density, and correctly characterizing the physics at the interface of the two repulsive species\cite{lous2018}.
\\

\section{Main theoretical results} 
\noindent 
We review the main theoretical studies on shell-shaped quantum gases and we discuss the physics of curved atomic waveguides {and rings. 
The former} works were largely inspired by the experiments in microgravity\cite{carollo2022}, and contributed to generate the broad perspective of studying quantum gases in curved geometries.
\\

\subsection{Shell-shaped quantum gases}
\noindent 
The properties of shell-shaped bosonic gases were extensively investigated in the last few years (see Box 1). 
Fundamental studies have discussed the phenomenon of Bose-Einstein condensation for a gas on the surface of a sphere\cite{tononi2019}, analyzing the interacting case with Bogoliubov theory\cite{tononi2019}, and both thin and thick spherical shells of noninteracting bosons were then discussed \cite{bereta2019}. 
Later on, the aim to model the experiments in microgravity \cite{carollo2022} led to consider the ellipsoidal geometry, determining the density distribution in the presence of trap inhomogeneities \cite{lundblad2019}, as well as the critical temperature and the free expansion of thin prolate shells\cite{tononi2020,rhyno2021}. 
Very recent papers considered a two-component Fermi gas on the sphere, and analyzed its crossover from a Bardeen-Cooper-Schrieffer regime to the BEC regime controlled by varying the radius\cite{he2022}, while other studies derived the equation of state of a spherical Bose gas\cite{tononi2022b} and its thermodynamics \cite{rhyno2021, tononi2022}. 

The clearest evidence of the curved geometry is perhaps provided by the dynamic properties of the shell. Among these, the long-wavelength collective modes reveal useful information on the physical regime of the system, and are typically easy to excite by perturbing a gas initially at rest. 
Remarkably, some zero-temperature collective modes show discontinuities along the transition from a filled spherical condensate to a thin shell\cite{sun2018,padavic2018,wolf2022}, and thus provide a clear way to experimentally observe the hollowing transition. 
{The zero}-temperature modes, therefore, were widely studied in the last years due to their importance \cite{lannert2007, diniz2020, moller2020, andriati2021}. 

Another direction of research on two-dimensional shell-shaped gases focused on transitions and on hydrodynamics at finite temperature, which must take into account that a normal, viscous, component coexists with the superfluid one.
The vanishing of the superfluid density due to the thermal proliferation of vorticity, described by the Berezinskii-Kosterlitz-Thouless transition (BKT), was recently derived for shell-shaped superfluids\cite{tononi2022}, by developing related works of the last decades \cite{kotsubo1984, ovrut1991,mitra2008} in view of the new experiments. 
In the hydrodynamic description of a two-dimensional flat superfluid, two sound modes can coexist and the second sound has been shown to vanish at the superfluid transition \cite{christodoulou2021}. 
The finite-temperature hydrodynamic modes of the spherical shells also display this behavior\cite{tononi2022}, and can then be used to check the superfluidity of a quantum gas with the topology of a sphere. 
\\

\vspace{8mm}
\noindent
\fbox{%
\parbox{0.48\textwidth}{%
Box 1 \textbar \ \textbf{Shell-shaped bosonic gases.} 
\begin{itemize}
\item Experimentally realized with radiofrequency-induced adiabatic potentials\cite{carollo2022} (FIG. \ref{fig2}a) and with repulsive Bose-Bose mixtures\cite{jia2022} (FIG. \ref{fig2}d).
These different platforms display complementary advantages and issues. The former allows to produce a 2D hollow shell but requires a microgravity environment\cite{lundblad2019}; the latter allows to tune the interatomic interactions and can be observed in ground-based experiments, but the shell (made of one species) must be filled by the other species.
\item The free expansion of the shell shows the self interference of the Bose-Einstein condensate\cite{jia2022,tononi2020}, a feature which is {reproducible in different experimental shots} due to the system coherence, and which could ease the imaging of vortices\cite{caracanhas2022}. 
\item The collective modes of the shell should show a distinctive non-monotonic behavior at the hollowing transition\cite{sun2018,padavic2018}, when the gas is inflated from a harmonic condensate to a bubble, and at the 2D superfluid transition\cite{tononi2022}, when the quantized vortices proliferate and prevent superfluidity.
\item Future prospects include studying the dynamics of quantized vortices in 2D curved superfluids \cite{bereta2021,caracanhas2022} and the analog simulation of cosmological phenomena such as inflation\cite{eckel2018}, the rapid expansion of the early universe, with expanding superfluid shells.
\end{itemize}
}%
}
\\

\subsection{Curved waveguides} 
\noindent
{ The Schr\"odinger equation for a particle constrained on a curved manifold has been analyzed in several papers\cite{switkes1977, dacosta1981, dacosta1982, kaplan1997} since the 1970s. 
Relatively recent investigations, then,} have discussed the role of curvature within the nonlinear Schr\"odinger equation, i.e. the Gross-Pitaevskii equation, to describe quasi-1D Bose-Einstein condensates confined in curved waveguides\cite{leboeuf2001, cozzini2006, sandin2017, salasnich2022, delcampo2014}. 
These papers show that the 3D equation of the constrained system can be reduced to an effective 1D equation with an additional potential, $U(s)=\hbar^2\kappa(s)^2/(8m)$, that depends on the local curvature $\kappa(s)$, where $m$ is the atomic mass and $s$ is the local (arclength) coordinate.
For instance, in the case of a Bose-Einstein condensate in an elliptical waveguide, $U(s)$ has the shape of a double-well potential, whose depth depends on the eccentricity of the ellipse \cite{salasnich2022}. In this system, the mean-field density of the ground state displays a transition from a two-peak configuration to a single-peak configuration at a critical attractive interaction strength \cite{salasnich2022}. 

The theoretical study of 1D waveguides with locally-varying curvature is still underdeveloped, partly because the geometric potential provides a small contribution for the typical experimental parameters. 
Indeed, for a local curvature in the regime $\kappa \sim \mu\text{m}^{-1}$, the geometric potential is $U/h \sim 10 \, \text{Hz}$, which is much lower than the typical chemical potential $\mu/h \sim [10^2;10^3] \, \text{Hz}$ or of typical trap inhomogeneities\cite{lundblad2019}. 
When neglecting the geometric potential, the boundary condition is the only feature characterizing the geometry of a ring-shaped system as curved (FIG.~\ref{fig1}b). 
Along this line, there were various studies, broadly connected to the emerging field of atomtronics\cite{amico2022}, on persistent currents and their damping in rings\cite{cominotti2014,polo2018,polo2019}, or on the investigation of interferometric properties with spinor condensates \cite{helm2018}. 

\section{Theoretical prospects}
\noindent 
Several recent papers analyzed the physics of quantum gases in curved geometries, aiming at different research questions and adopting a variety of techniques. 
We propose a unified framing and categorize them into a few research directions. 
\\

\subsection{Exploring different geometries}
\noindent
In the last years, the analyses traditionally performed in flat gases have been extended in the context of spherical shells, investigating the stability of quantum mixtures \cite{andriati2021}, the role of dipolar interactions \cite{diniz2020,arazo2021}, and the excitation spectrum of a slightly prolate spherical gas\cite{biral2022}. 
Much less attention, however, has been dedicated to generic manifolds, whose study would allow to obtain general results, and to specific geometries such as ellipsoids, tori, cylinders, M\"obius strips, and 1D bent waveguides. 

Further investigations could then search for the geometric corrections, controlled by the system size and by the density of states, to the equation of state and to the phase transitions of these geometries. 
For instance, the BKT transition of a superfluid confined on the cylinder surface should be greatly modified\cite{machta1989}, since the interaction energy of the vortices scales linearly with distance, and the vortex proliferation that induces the vanishing of superfluidity should not occur. 
In a similar way, when increasing the aspect ratio in {the} ellipsoidal and in {the} toroidal geometries, the BKT transition should {smoothly} crossover from the usual jump of the superfluid density into a {thermally-driven superfluid} transition that occurs in the absence of renormalization. 

In the various geometries, the analysis of the condensate dynamics could focus on the collective modes, whose decay could be suppressed by symmetries that are preserved, and therefore protected, by the system topology\cite{li2022}. 
In one-dimensional superfluid circuits, the role of curvature on persistent currents and their eventual heating induced by the curved path could be investigated. 
Finally, the realization of vortex lattices on curved manifolds could realize lattice models such as the Hofstadter-Hubbard model on a cylinder \cite{lacki2016}, and could provide an alternative way to synthetic dimensions for simulating the quantum Hall physics in cylindrical geometry \cite{hsunli2022}. 
\\

\begin{figure*}[hbtp]
\resizebox{0.99 \textwidth}{!}{\includegraphics*{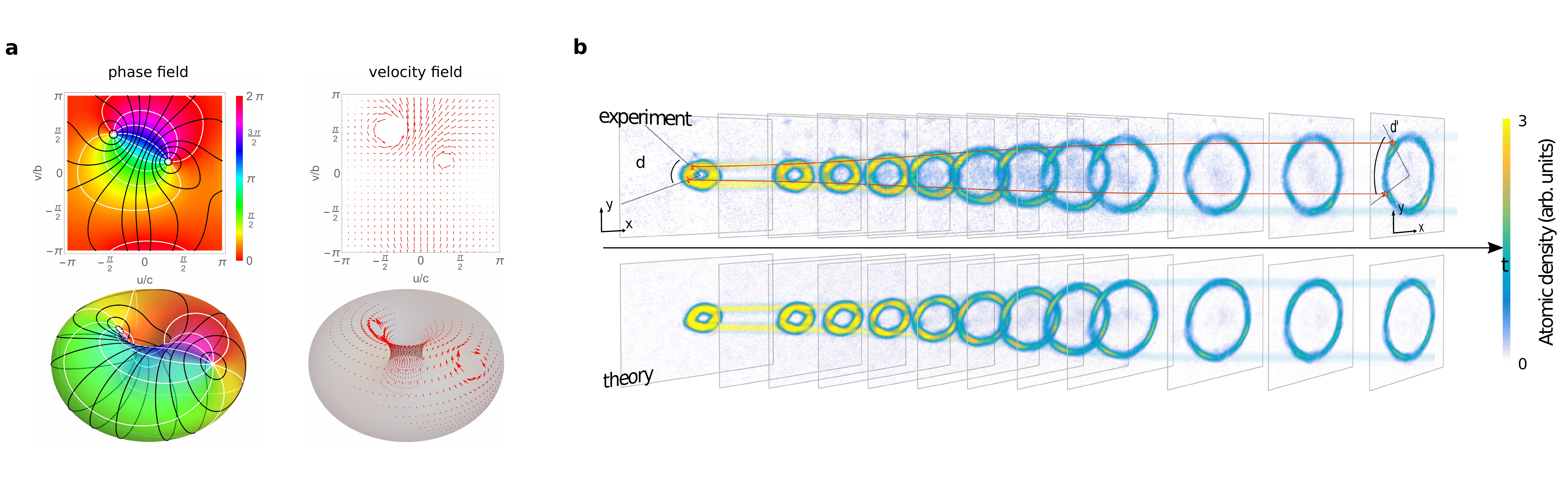}}
\caption{{\bf Prospects.} Prospects of future research may involve analyzing the quantum statistics and the dynamics of quantum gases confined in different geometries.
\textbf{a} \textbar \ A compact curved surface imposes topological constraints on the phase of the macroscopic wave function, and thus generate superfluid currents and motion of vortex-antivortex pairs that provide a different signature to each topology and geometry. This example depicts the phase field and the superfluid velocity field of a vortex-antivortex dipole moving on the surface of a torus\cite{guenther2020}. 
\textbf{b} \textbar \ By mapping the time evolution of a quantum gas in a curved ring geometry, the experiment of ref.\cite{eckel2018} realized an analog quantum simulator of cosmological physics. 
The supersonic expansion of the superfluid allowed to simulate both the redshift of photons and the Hubble friction.
Part \textbf{a}
adapted with permission from ref.\cite{guenther2020}, APS.
Part \textbf{b}
adapted with permission from ref.\cite{eckel2018}, APS.
\\
} \label{fig3}
\end{figure*}

\subsection{Vortices and solitons}
\noindent
The physics of vortices in two-dimensional curved geometries has been studied both in classical fluids \cite{newton2001} and in superfluids \cite{turner2010}, since they connect the in-plane fluid motion with the curvature of the underlying manifold.
The usual models developed in zero-temperature superfluids assume an irrotational and incompressible fluid, thus neglecting the density fluctuations and the local variation of the transverse confinement, so that vortices are point-like defects with integer topological charge $q$, and the superfluid density $n_s$ is uniform across the surface. 
Under these hypotheses, the motion of $N$ vortices is determined by the emergent system Hamiltonian\cite{turner2010},
\begin{equation*}
\frac{H}{K_s} = \sum_{i<j}^N 4 \pi^2 q_i  q_j \, V_{ij}(\mathbf{r}_i,\mathbf{r}_j) - \sum_i^N \pi q_i^2 U_G(\mathbf{r}_i) ,  
\end{equation*}
with $K_s=\hbar^2 n_s/m$ the superfluid stiffness for a system of bosonic atoms with mass $m$. 
In particular, the adimensional vortex-vortex interaction $V_{ij}(\mathbf{r}_i,\mathbf{r}_j)$ is the Green's function of the covariant Laplacian [which reduces to $(2\pi)^{-1}\ln{(|\mathbf{r}_i-\mathbf{r}_j|/\xi)}$ in the flat case, with $\xi$ the vortex core size]. 
The geometric potential $U_G$ is determined by the system curvature, since it satisfies the covariant Poisson equation with the Gaussian curvature $K(\mathbf{r})$ as a source, $\nabla^2 U_G(\mathbf{r})=K(\mathbf{r})$. 
Remarkably, the single-valuedness of the phase field of the superfluid plays a key role in determining the geometric potential and the form of the two-vortex interaction, and it is therefore the source of geometry-dependent vortex motion, as shown for instance by the analysis of cylindric superfluids \cite{guenther2017}. 

In the recent years, studies of vortex dynamics were performed in a large variety of two-dimensional superfluid manifolds, which include the cylinder \cite{guenther2017}, the cone \cite{massignan2019}, the torus \cite{guenther2020,ambroise2022}, the sphere \cite{padavic2020,bereta2021}, and ellipsoids or other generic axially-symmetric surfaces \cite{caracanhas2022}. 
In all these cases, the system curvature and the boundary conditions are the source of peculiar dynamical effects which are intrinsically characteristic of each realization. 
A general important aspect is that the vortices in a uniform superfluid couple with surface curvature, and thus feel a geometric potential which repels them from positive curvature regions and attracts them towards negative curvature ones \cite{vitelli2004}.

A series of developments of these analyses could be imagined: while an analytical framework to integrate the vortex motion is available for a uniform and incompressible superfluid\cite{turner2010}, the relaxation of these hypotheses may generate interesting phenomena. 
In specific regimes, the geometric potential could compete with the other energy contributions that appear when, for instance, the superfluid density $n_s$ is nonuniform, and when the vortex core size and the energy density can change due to density fluctuations or due to the surface curvature. 
A general framework to describe analytically this case is currently absent, and the modeling of vortex motion which includes these and, potentially, also the temperature-dependent effects requires numerical simulations \cite{blakie2008,proukakis2008}. 
Other research directions include the study of spinor superfluids in various geometries, following the interesting evidence that two types of vortices with the same topological charge can exist in the cylindrical geometry\cite{ho2015}. 
Finally, in the case of Bose-Einstein condensates in an elliptical ring, we expect that multi-quantized vortices, and in particular the position of their cores, will strongly depend on the eccentricity, which determines the local curvature. Also the stability of vortex configurations may be affected by the eccentricity, and this phenomenon could be then used to build sensors based on the shape deformation of the elliptical waveguide.

The investigation of nonlinear superfluid excitations should also include bright and dark solitons in bosonic gases with attractive { or repulsive} interactions, which can be studied both in two-dimensional curved settings and in one-dimensional waveguides.
For instance, it was proposed to realize a matter-wave interferometer in a curved ring geometry by splitting a bright soliton into two counterpropagating components along the ring\cite{helm2015}.
In general, it is still unclear how bright solitons in attractive gases, dark solitons in repulsive condensates, and multi-soliton configurations could exist (and if they are stable) if their motion is constrained on a curved low-dimensional geometry. 
Investigating their stability, as well as their scattering and the finite-temperature effects on various curved manifolds will pose new challenges both for theory and for experiments. 
\\

\subsection{Few-body physics} 
\noindent
While many-body analyses of curved quantum systems could be technically complicated by the necessity of using involved coordinate systems, focusing on the few-body properties could facilitate the task. 
A relatively limited number of two-body problems has been already solved in the spherical geometry, analyzing the scattering of particles with zero-range interactions in all partial waves \cite{zhang2018,tononi2022}, and the energy spectrum of anyons on the sphere \cite{ouvry2019}. 
The interesting physics in the study of scattering problems, resonances and binding, should occur when the energy associated to the system curvature is comparable to that of the phenomenon considered. 

Detailed investigations should be specifically focused on understanding the interplay between the spatial curvature and the bound states. 
For instance, chiral bound states on the surface of a sphere feel the vector potential of an magnetic monopole located at the origin\cite{shi2017}. 
This phenomenon, which is the effective consequence of the motion on a curved surface, should be further studied, since it could have implications for new many-body physics and be at the basis of topological chiral currents in the system. 
\\

\subsection{Analog simulation and mappings}
\noindent
Analog quantum simulation aims to simulate phenomena that are unaccessible to experiments or to theory: from condensed matter to high energy physics, many systems can be studied by mapping their Hamiltonian to that of controllable setups as, for instance, ultracold atoms experiments \cite{bloch2012,georgescu2014}. 
In low-dimensional Bose-Einstein condensates, most simulations rely on some mapping between the measurable dynamic properties of the condensate and the metric tensor of a curved gravitational system \cite{hu2019,munozdenova2019,viermann2022}. 
However, even if there have been studies of analog models based on BECs in a curved geometry\cite{visser2001}, the curvature itself as a tunable degree of freedom of the simulation remains a largely unexplored feature. 

{A few experiments with ring-shaped condensates have simulated various} cosmological phenomena, analyzing the expansion and subsequent preheating of the universe \cite{eckel2018}, measuring the Hubble attenuation and amplification in expanding and contracting rings \cite{banik2022}, and a few theoretical proposals were elaborated \cite{gomezllorente2019,bhardwaj2021,eckel2021}. 
The possibilities of building analog models go surely beyond these notable realizations, and should be thoroughly considered in two-dimensional curved settings. 

Finally, there is a fundamental correspondence between the physics of rapidly rotating superfluids, which develop vortex lattices, and the quantum Hall physics\cite{ho2001,cooper2008,fetter2009}, which occurs when a two-dimensional electron gas is placed in a strong magnetic field. 
Studies of rapidly rotating superfluids in the flat case \cite{ho2001,mukherjee2022,schweikhard2004} could be extended in various geometries, as well as investigating vortices in spinor condensates in various geometries \cite{ho2015}. 
\\

\section{Conclusions}
\noindent
Many research areas in physics, such as magnetism \cite{streubel2016}, electronics \cite{gentile2022}, graphene\cite{guinea2010}, and liquid crystals \cite{turner2010} are extending their low-dimensional analyses into the third dimension, searching for new phenomena induced by the curved geometry. 
In systems of ultracold atoms, the prospect of having new configurations besides shell-shaped and ring-shaped gases is inspiring further theoretical developments that can advance the study of many-body physics of curved quantum systems. 

In this Perspective, we analyzed the most promising research topics in low-dimensional curved quantum gases, reviewing the research done on vortices, on the few-body physics and on quantum simulation. 
We expect that exploring the interplay of curvature, topology and geometry in various curved configurations could both lead to future technological applications and to fundamental scientific advances. 

\vspace{0mm}
\begin{acknowledgements}
A.~T.~acknowledges support from ANR grant ``Droplets'' No.~ANR-19-CE30-0003-02. 
L.~S.~is partially supported by the BIRD grant ``Ultracold atoms in curved geometries'' of the University of Padova, by the ``Iniziativa Specifica Quantum'' of INFN, {and by  the European Union-NextGenerationEU within the National Center for
HPC, Big Data and Quantum Computing (Project No. CN00000013, CN1 Spoke 1: “Quantum Computing”)}. A.~T.~thanks R.~Dubessy and I.~B.~Spielman for interesting discussions. 
{ L.S. thanks A. Yakimenko for useful suggestions.}
\end{acknowledgements}

\vspace{4mm}
\begin{center}
\textbf{\small Competing interests}

\vspace{4mm}
The authors declare no competing interests.
\end{center}


\begin{thebibliography}{100}

\bibitem{carollo2022} Carollo, R. A. et al. Observation of ultracold atomic bubbles in orbital microgravity. \textit{Nature} \textbf{606}, 281-286 (2022).

\bibitem{jia2022} Jia, F., Huang, Z., Qiu, L., Zhou, R., Yan, Y. \& Wang, D. Expansion Dynamics of a Shell-Shaped Bose-Einstein Condensate. \textit{Phys. Rev. Lett.} \textbf{129}, 243402 (2022).

\bibitem{amico2022} Amico, L., et al. Colloquium: Atomtronic circuits: From many-body physics to quantum technologies. \textit{Rev. Mod. Phys.} \textbf{94}, 041001 (2022).

\bibitem{navon2021} Navon, N., Smith, R. P. \& Hadzibabic, Z. Quantum gases in optical boxes. \textit{Nat. Phys.} \textbf{17}, 1334–1341 (2021).

\bibitem{dalfovo1999} Dalfovo, F. Giorgini, S., Pitaevskii, L. P. \& Stringari, S. Theory of Bose-Einstein condensation in trapped gases. \textit{Rev. Mod. Phys.} \textbf{71}, 463 (1999).

\bibitem{tononi2019} Tononi, A. \& Salasnich, L. Bose-Einstein Condensation on the Surface of a Sphere. \textit{Phys. Rev. Lett.} \textbf{123}, 160403 (2019).

\bibitem{tononi2022} Tononi, A., Pelster, A. \& Salasnich, L.
Topological superfluid transition in bubble-trapped condensates.
\textit{Phys. Rev. Res.} \textbf{4}, 013122 (2022).

\bibitem{lannert2007} Lannert, C., Wei, T.-C. \& Vishveshwara, S. Dynamics of condensate shells: Collective modes and expansion. \textit{Phys. Rev. A} \textbf{75}, 013611 (2007).

\bibitem{sun2018} Sun, K., Padavi\'c, K., Yang, F., Vishveshwara, S. \& Lannert, C. Static and dynamic properties of shell-shaped condensates. \textit{Phys. Rev. A} \textbf{98}, 013609 (2018).

\bibitem{padavic2018} Padavi\'c, K., Sun, K., Lannert, C. \& Vishveshwara, S. Physics of hollow Bose-Einstein condensates. \textit{EPL} \textbf{120}, 20004 (2018).

\bibitem{guo2022} Guo, Y., Mercado Gutierrez, E., Rey, D., Badr, T., Perrin, A., Longchambon, L., Bagnato, V. S., Perrin, H. \& Dubessy, R. Expansion of a quantum gas in a shell trap. \textit{New J. Phys.} \textbf{24}, 093040 (2022).

\bibitem{rey2022} Rey, D., Thomas, S., Sharma, R., Badr, T., Longchambon, L., Dubessy, R. \& Perrin, H. Loading a quantum gas from an hybrid dimple trap to a shell trap. Preprint at https://arxiv.org/abs/2208.14684 (2022).

\bibitem{guenther2017} Guenther, N.-E., Massignan, P. \& Fetter, A. L. Quantized superfluid vortex dynamics on cylindrical surfaces and planar annuli. \textit{Phys. Rev. A} \textbf{96}, 063608 (2017).

\bibitem{caracanhas2022} Caracanhas, M. A., Massignan, P. \& Fetter, A. L. Superfluid vortex dynamics on an ellipsoid and other surfaces of revolution. \textit{Phys. Rev. A} \textbf{105}, 023307 (2022).

\bibitem{guenther2020} Guenther, N.-E., Massignan, P. \& Fetter, A. L. Superfluid vortex dynamics on a torus and other toroidal surfaces of revolution. \textit{Phys. Rev. A} \textbf{101}, 053606 (2020).

\bibitem{eckel2018} Eckel, S., Kumar, A., Jacobson, T., Spielman, I. B. \& Campbell, G. K. A Rapidly Expanding Bose-Einstein Condensate: An Expanding Universe in the Lab. \textit{Phys. Rev. X} \textbf{8}, 021021 (2018).

\bibitem{guo2020} Guo, Y., Dubessy, R., de Go\"{e}r de Herve, M., Kumar, A., Badr, T., Perrin, A., Longchambon, L. \& Perrin, H. Supersonic Rotation of a Superfluid: A Long-Lived Dynamical Ring. \textit{Phys. Rev. Lett.} \textbf{124}, 025301 (2020).

\bibitem{pitaevskii2016} Pitaevskii, L. P. \& Stringari, S. \textit{Bose-Einstein Condensation and Superfluidity} (Oxford Univ. Press, 2016).

\bibitem{dewitt1957} DeWitt, B. S. Dynamical Theory in Curved Spaces. I. A Review of the Classical and Quantum Action Principles. \textit{Rev. Mod. Phys.} \textbf{29}, 377 (1957).

\bibitem{dacosta1982} Da Costa, R. C. T. Constraints in quantum mechanics. \textit{Phys. Rev. A} \textbf{25}, 2893 (1982).

\bibitem{jost2010} Jost, J. \textit{Riemannian Geometry and Geometric Analysis} (Springer, 2010).

\bibitem{leboeuf2001} Leboeuf, P. \& Pavloff, N. Bose-Einstein beams: Coherent propagation through a guide. \textit{Phys. Rev. A} \textbf{64}, 033602 (2001).

\bibitem{cozzini2006} Schwartz, S., Cozzini, M., Menotti, C., Carusotto, I., Bouyer, P. \& Stringari, S. One-dimensional description of a Bose-Einstein condensate in a rotating closed-loop waveguide. \textit{New J. Phys.} \textbf{8}, 162 (2006).

\bibitem{sandin2017} Sandin, P., Ogren, M., Gulliksson, M., Smyrnakis, J., Magiropoulos, M. \& Kavoulakis, G. M. Dimensional reduction in Bose-Einstein condensed clouds of atoms confined in tight potentials of any geometry and any interaction strength. \textit{Phys. Rev. E} \textbf{95}, 012142 (2017).

\bibitem{salasnich2022} Salasnich, L. Bose-Einstein condensate in an elliptical waveguide. \textit{SciPost Phys. Core} \textbf{5}, 015 (2022). 

\bibitem{boada2012} Boada, O., Celi, A., Latorre, J. I. \& Lewenstein, M. Quantum Simulation of an Extra Dimension. \textit{Phys. Rev. Lett.} \textbf{108}, 133001 (2012).

\bibitem{boada2015} Boada, O., Celi, A., Rodríguez-Laguna, J., Latorre, J. I. \& Lewenstein, M. Quantum simulation of non-trivial topology. \textit{New J. Phys.} \textbf{17}, 045007 (2015).

\bibitem{ozawa2019} Ozawa, T., Price, H. M. Topological quantum matter in synthetic dimensions. \textit{Nat. Rev. Phys.} \textbf{1}, 349–357 (2019).

\bibitem{zobay2000} Zobay, O. \& Garraway, B. M. Properties of coherent matter-wave bubbles. \textit{Acta physica slovaca} \textbf{2000}, 3 (2000).

\bibitem{zobay2001} Zobay, O. \& Garraway, B. M. Two-Dimensional Atom Trapping in Field-Induced Adiabatic Potentials. \textit{Phys. Rev. Lett.} \textbf{86}, 1195 (2001).

\bibitem{zobay2004} Zobay, O. \& Garraway, B. M. Atom trapping and two-dimensional Bose-Einstein condensates in field-induced adiabatic potentials. \textit{Phys. Rev. A} \textbf{69}, 023605 (2004).

\bibitem{colombe2004} Colombe, Y. et al. Ultracold atoms confined in rf-induced two-dimensional trapping potentials. \textit{EPL} \textbf{67}, 593 (2004).

\bibitem{white2006} White, M., Gao, H., Pasienski, M. \& DeMarco, B. Bose-Einstein condensates in rf-dressed adiabatic potentials. \textit{Phys. Rev. A} \textbf{74}, 023616 (2006).

\bibitem{merloti2013} Merloti, K. et al. A two-dimensional quantum gas in a magnetic trap.  \textit{New J. Phys.} \textbf{15}, 033007 (2013).

\bibitem{harte2018} Harte, T. L. et al. Ultracold atoms in multiple radio-frequency dressed adiabatic potentials. \textit{Phys. Rev. A} \textbf{97}, 013616 (2018).

\bibitem{lundblad2019} Lundblad, N. et al. Shell potentials for microgravity Bose-Einstein condensates. \textit{npj Microgravity} \textbf{5}, 30 (2019).

\bibitem{elliott2018} Elliott, E. R., Krutzik, M. C., Williams, J. R., Thompson, J. R. \& Aveline, D. C. NASA's Cold Atom Lab (CAL): system development and ground test status. \textit{npj Microgravity} \textbf{4}, 16 (2018).

\bibitem{aveline2020} Aveline, D. C. et al. Observation of Bose-Einstein condensates in an Earth-orbiting research lab. \textit{Nature} \textbf{582}, 193-197 (2020).

\bibitem{luksch2019} Luksch, K., et al. Probing multiple-frequency atom-photon interactions with ultracold atoms. \textit{New J. Phys.} \textbf{21}, 073067 (2019).

\bibitem{fernholz2007} Fernholz, T., Gerritsma, R., Kr\"{u}ger, P. \& Spreeuw, R. J. C. Dynamically controlled toroidal and ring-shaped magnetic traps. \textit{Phys. Rev. A} \textbf{75}, 063406 (2007).

\bibitem{lesanovsky2007} Lesanovsky, I. \& von Klitzing, W. Time-Averaged Adiabatic Potentials: Versatile Matter-Wave Guides and Atom Traps. \textit{Phys. Rev. Lett.} \textbf{99}, 083001 (2007).

\bibitem{sherlock2011} Sherlock, B. E., Gildemeister, M., Owen, E., Nugent, E. \& Foot, C. J. Time-averaged adiabatic ring potential for ultracold atoms. \textit{Phys. Rev. A} 83, 043408 (2011); \textit{Phys. Rev. A} \textbf{83}, 059904 (2011).

\bibitem{amico2021} Amico, L. et al. Roadmap on Atomtronics: State of the art and perspective featured. \textit{AVS Quantum Sci.} \textbf{3}, 039201 (2021).

\bibitem{garraway2016} Garraway, B. M. \& Perrin, H. Recent developments in trapping and manipulation of atoms with adiabatic potentials. \textit{J. Phys. B: At. Mol. Opt. Phys.} \textbf{49}, 172001 (2016).

\bibitem{gupta2005} Gupta, S., Murch, K. W., Moore, K. L., Purdy, T. P. \& Stamper-Kurn, D. M. Bose-Einstein Condensation in a Circular Waveguide. \textit{Phys. Rev. Lett.} \textbf{95}, 143201 (2005).

\bibitem{ryu2007} Ryu, C. et al. Observation of Persistent Flow of a Bose-Einstein Condensate in a Toroidal Trap. \textit{Phys. Rev. Lett.} \textbf{99}, 260401 (2007).

\bibitem{ramanathan2011} Ramanathan, A. et al. Superflow in a Toroidal Bose-Einstein Condensate: An Atom Circuit with a Tunable Weak Link. \textit{Phys. Rev. Lett.} \textbf{106}, 130401 (2011). 

\bibitem{eckel2014} Eckel, S., Jendrzejewski, F., Kumar, A., Lobb, C. J. \& Campbell, G. K. Interferometric Measurement of the Current-Phase Relationship of a Superfluid Weak Link. \textit{Phys. Rev. X} \textbf{4}, 031052 (2014). 

\bibitem{eckel2014b} Eckel, W. et al. Hysteresis in a quantized superfluid ‘atomtronic’ circuit. \textit{Nature} \textbf{506}, 200-203 (2014). 

\bibitem{ryu2015} Ryu, C. \& Boshier, M. G. Integrated coherent matter wave circuits. \textit{New J. Phys.} \textbf{17}, 092002 (2015).

\bibitem{pandey2019} Pandey, S. et al. Hypersonic Bose--Einstein condensates in accelerator rings. \textit{Nature} \textbf{570}, 205–209 (2019).

\bibitem{degoerdeherve2021} de Go\"{e}r de Herve, M. et al. A versatile ring trap for quantum gases. \textit{J. Phys. B: At. Mol. Opt. Phys.} \textbf{54}, 125302 (2021).

\bibitem{cai2022} Cai, Y., Allman, D. G., Sabharwal, P. \& Wright, K. C. Persistent Currents in Rings of Ultracold Fermionic Atoms. \textit{Phys. Rev. Lett.} \textbf{128}, 150401 (2022). 

\bibitem{moulder2012}
Moulder, S., Beattie, S., Smith, R. P., Tammuz, N. \& Hadzibabic, Z. Observation of persistent flow of a Bose-Einstein condensate in a toroidal trap, \textit{Phys. Rev. A} \textbf{86}, 013629 (2012). 

\bibitem{dubessy2012} Dubessy, R., Liennard, T., Pedri, P. \& Perrin, H. Critical rotation of an annular superfluid Bose-Einstein condensate, \textit{Phys. Rev. A} \textbf{86}, 011602(R) (2012). 

\bibitem{sinucoleon2014} Sinuco-Le\'{o}n, G. et al. Inductively guided circuits for ultracold dressed atoms. \textit{Nat. Commun.} \textbf{5}, 5289 (2014).

\bibitem{ho1996} Ho, T.-L. \& Shenoy, V. B. Binary Mixtures of Bose Condensates of Alkali Atoms. \textit{Phys. Rev. Lett.} \textbf{77}, 3276 (1996).

\bibitem{pu1998} Pu, H. \& Bigelow, N. P. Properties of Two-Species Bose Condensates. \textit{Phys. Rev. Lett.} \textbf{80}, 1130 (1998).

\bibitem{wolf2022} Wolf, A. et al. Shell-shaped Bose-Einstein condensates based on dual-species mixtures. \textit{Phys. Rev. A} \textbf{106}, 013309 (2022).

\bibitem{derrico2019} D'Errico, C. et al. Observation of quantum droplets in a heteronuclear bosonic mixture. \textit{Phys. Rev. Res.} \textbf{1}, 033155 (2019).

\bibitem{ospelkaus2006} Ospelkaus-Schwarzer, S. Quantum degenerate fermi-bose mixtures of 40K and 87Rb in 3D optical lattices. (Universit\"{a}t Hamburg, 2006).

\bibitem{safronova2006} Safronova, M. S., Arora, B. \& Clark, C. W. Frequency-dependent polarizabilities of alkali-metal atoms from ultraviolet through infrared spectral regions. \textit{Phys. Rev. A} \textbf{73}, 1 (2006).

\bibitem{onofrio2002}
Onofrio R. \& Presilla, C. Reaching Fermi Degeneracy in Two-Species Optical Dipole Traps. \textit{Phys. Rev. Lett.} \textbf{89}, 100401 (2002).

\bibitem{frye2021} Frye, K. et al. The Bose-Einstein Condensate and Cold Atom Laboratory. \textit{EPJ Quantum Technol.} \textbf{8}, 1 (2021).

\bibitem{thompson2023} Thompson, R. J. et al. Exploring the quantum world with a third generation ultra-cold atom facility. \textit{Quantum Sci. Technol.} \textbf{8}, 014007 (2023).

\bibitem{vanzoest2010} van Zoest, T. et al. Bose-Einstein Condensation in Microgravity. \textit{Science} \textbf{328}, 1540-1543 (2010).

\bibitem{condon2019} Condon, G. et al. All-Optical Bose-Einstein Condensates in Microgravity. \textit{Phys. Rev. Lett.} \textbf{123}, 240402 (2019).

\bibitem{wille2008} Wille, E. et al. Exploring an Ultracold Fermi-Fermi Mixture: Interspecies Feshbach Resonances and Scattering Properties of $^6$Li and $^{40}$K. \textit{Phys. Rev. Lett.} \textbf{100}, 053201 (2008).

\bibitem{voigt2009} Voigt, A.-C. et al. Ultracold Heteronuclear Fermi-Fermi Molecules. \textit{Phys. Rev. Lett.} \textbf{102}, 020405 (2009).

\bibitem{hara2011} Hara, H., Takasu, Y., Yamaoka, Y., Doyle, J. M. \& Takahashi, Y. Quantum Degenerate Mixtures of Alkali and Alkaline-Earth-Like Atoms. \textit{Phys. Rev. Lett.} \textbf{106}, 205304 (2011).

\bibitem{ferrierbarbut2014} Ferrier-Barbut, I. et al. A mixture of Bose and Fermi superfluids. \textit{Science} \textbf{345}, 1035-1038, (2014).

\bibitem{wang2016} Wang, F., Li, X., Xiong, D. \& Wang, D. A double species $^{23}$Na and $^{87}$Rb Bose--Einstein condensate with tunable miscibility via an interspecies Feshbach resonance. \textit{J. Phys. B: At. Mol. Opt. Phys.} \textbf{49}, 015302 (2016).

\bibitem{roy2017} Roy, R., Green, A., Bowler, R. \& Gupta, S. Two-Element Mixture of Bose and Fermi Superfluids. \textit{Phys. Rev. Lett.} \textbf{118}, 055301 (2017).

\bibitem{cabrera2018} Cabrera, C. R. et al. Quantum liquid droplets in a mixture of Bose-Einstein condensates. \textit{Science} \textbf{359}, 301 (2018).

\bibitem{burchianti2018} Burchianti, A. et al. Dual-species Bose-Einstein condensate of $^{41}$K and $^{87}$Rb in a hybrid trap. \textit{Phys. Rev. A} \textbf{98}, 063616 (2018).

\bibitem{semeghini2018} Semeghini, G. et al. Self-Bound Quantum Droplets of Atomic Mixtures in Free Space. \textit{Phys. Rev. Lett.} \textbf{120}, 235301 (2018).

\bibitem{neri2020} Neri, E. et al. Realization of a cold mixture of fermionic chromium and lithium atoms. \textit{Phys. Rev. A} \textbf{101}, 063602 (2020).

\bibitem{green2020} Green, A. et al. Feshbach Resonances in p-Wave Three-Body Recombination within Fermi-Fermi Mixtures of Open-Shell $^6$Li and Closed-Shell $^{173}$Yb Atoms. \textit{Phys. Rev. X} \textbf{10}, 031037 (2020).

\bibitem{ravensbergen2018} Ravensbergen, C. et al. Production of a degenerate Fermi-Fermi mixture of dysprosium and potassium atoms. \textit{Phys. Rev. A} \textbf{98}, 063624 (2018).

\bibitem{lous2018} Lous, R. S. et al. Probing the Interface of a Phase-Separated State in a Repulsive Bose-Fermi Mixture. \textit{Phys. Rev. Lett.} \textbf{120}, 243403 (2018).

\bibitem{bereta2019} Bereta, S. J., Madeira, L., Bagnato, V. S. \& Caracanhas, M. A. Bose--Einstein condensation in spherically symmetric traps. \textit{Am. J. Phys.} \textbf{87}, 924 (2019). 

\bibitem{tononi2020} Tononi, A., Cinti, F. \& Salasnich, L. Quantum bubbles in microgravity. \textit{Phys. Rev. Lett.} \textbf{125}, 010402 (2020).

\bibitem{rhyno2021} Rhyno, B., Lundblad, N., Aveline, D. C., Lannert, C. \& Vishveshwara, S. Thermodynamics in expanding shell-shaped Bose-Einstein condensates. \textit{Phys. Rev. A} \textbf{104}, 063310 (2021). 

\bibitem{he2022} He, Y., Guo, H. \& Chien, C.-C. BCS-BEC crossover of atomic Fermi superfluid in a spherical bubble trap. \textit{Phys. Rev. A} \textbf{105}, 33324 (2022). 

\bibitem{tononi2022b} Tononi, A. Scattering theory and equation of state of a spherical two-dimensional Bose gas. \textit{Phys. Rev. A} \textbf{105}, 023324 (2022).

\bibitem{diniz2020} Diniz, P. C. et al. Ground state and collective excitations of a dipolar Bose-Einstein condensate in a bubble trap. \textit{Sci. Rep.} \textbf{10}, 4831 (2020).

\bibitem{moller2020} M\'{o}ller, N. S., dos Santos, F. E. A., Bagnato, V. S. \& Pelster, A. Bose-Einstein condensation on curved manifolds. \textit{New J. Phys.} \textbf{22}, 063059 (2020).

\bibitem{andriati2021} Andriati, A., Brito, L., Tomio, L. \& Gammal, A. Stability of a Bose-condensed mixture on a bubble trap. \textit{Phys. Rev. A} \textbf{104}, 033318 (2021).

\bibitem{kotsubo1984} Kotsubo, V. \& Williams, G. A. Kosterlitz-Thouless Superfluid Transition for Helium in Packed Powders. \textit{Phys. Rev. Lett.} \textbf{53}, 691 (1984).

\bibitem{ovrut1991} Ovrut B. A. \& Thomas, S. Theory of vortices and monopoles on a sphere. \textit{Phys. Rev. D} \textbf{43}, 1314 (1991).

\bibitem{mitra2008} Mitra, K., Williams, C. J. \& Sa de Melo, C. A. R. Superfluid and Mott-insulating shells of bosons in harmonically confined optical lattices. \textit{Phys. Rev. A} \textbf{77}, 033607 (2008).

\bibitem{christodoulou2021} Christodoulou, P. et al. Observation of first and second sound in a BKT superfluid. \textit{Nature} \textbf{594}, 191-194 (2021).

\bibitem{switkes1977} Switkes, E., Russel, E. L. \& Skinner, J. L. Kinetic energy and path curvature in bound state systems. \textit{J. Chem. Phys.} \textbf{67}, 3061 (1977).

\bibitem{kaplan1997} Kaplan, L., Maitra, N. T. \& Heller, E. J. Quantizing constrained systems. \textit{Phys. Rev. A} \textbf{56}, 2592 (1997).

\bibitem{dacosta1981} da Costa, R. C. T. Quantum mechanics of a constrained particle. \textit{Phys. Rev. A} \textbf{23}, 1982 (1981).

\bibitem{delcampo2014} del Campo, A., Boshier, M. G. \& Saxena, A. Bent waveguides for matter-waves: supersymmetric potentials and reflectionless geometries. \textit{Sci. Rep.} \textbf{4}, 5274 (2014).

\bibitem{cominotti2014} Cominotti, M., Rossini, D., Rizzi, M., Hekking, F. \& Minguzzi, A. 
Optimal Persistent Currents for Interacting Bosons on a Ring with a Gauge Field. \textit{Phys. Rev. Lett.} \textbf{113}, 025301 (2014). 

\bibitem{polo2018} Polo, J., Ahufinger, V., Hekking, F. W. J. \& Minguzzi, A. Damping of Josephson Oscillations in Strongly Correlated One-Dimensional Atomic Gases. \textit{Phys. Rev. Lett.} \textbf{121}, 090404 (2018).

\bibitem{polo2019} Polo, J., Dubessy, R., Pedri, P., Perrin, H. \& Minguzzi, A. Oscillations and Decay of Superfluid Currents in a One-Dimensional Bose Gas on a Ring. \textit{Phys. Rev. Lett.} \textbf{123}, 195301 (2019).

\bibitem{helm2018} Helm, J. L., Billam, T. P., Rakonjac, A., Cornish, S. L. \& Gardiner, S. A.
Spin-orbit coupled interferometry with ring-trapped Bose--Einstein condensates, \textit{Phys. Rev. Lett.} \textbf{120}, 063201 (2018).

\bibitem{arazo2021} Arazo, M., Mayol, R. \& Guilleumas, M. Shell-shaped condensates with gravitational sag: contact and dipolar interactions. \textit{New J. Phys.} \textbf{23}, 113040 (2021).

\bibitem{biral2022} Biral, E. J. P., M\'{o}ller, N. S., Pelster, A. \& dos Santos, F. E. A. Bose-Einstein condensates and the thin-shell limit in anisotropic bubble traps. Preprint at https://arxiv.org/abs/2210.08074 (2022).

\bibitem{machta1989} Machta, J. \& Guyer, R. Superfluid films on a cylindrical surface. \textit{J. Low Temp. Phys.} \textbf{74}, 231 (1989).

\bibitem{li2022} Li, G. \& Efimkin, D. K. Equatorial Waves in Rotating Bubble-Trapped Superfluids. \textit{Phys. Rev. A} \textbf{107}, 023319 (2023).

\bibitem{lacki2016} {\L}acki, M. et al. Quantum Hall physics with cold atoms in cylindrical optical lattices. \textit{Phys. Rev. A} \textbf{93}, 013604 (2016).

\bibitem{hsunli2022} Li, C.-H. et al. Bose-Einstein Condensate on a Synthetic Topological Hall Cylinder. \textit{PRX Quantum} \textbf{3}, 010316 (2022).

\bibitem{newton2001} Newton, P. \textit{The N-Vortex Problem} (Springer-Verlag, New York, 2001).

\bibitem{turner2010} Turner, A. M., Vitelli, V. \& Nelson, D. R. Vortices on curved surfaces. \textit{Rev. Mod. Phys.} \textbf{82}, 1301 (2010).

\bibitem{massignan2019} Massignan, P. \& Fetter, A. L. Superfluid vortex dynamics on planar sectors and cones. \textit{Phys. Rev. A} \textbf{99}, 063602 (2019).

\bibitem{ambroise2022} D'Ambroise,  J., Carretero-Gonz\'{a}lez, R., Schmelcher, P. \& Kevrekidis, P. G. Superfluid vortex multipoles and soliton stripes on a torus. \textit{Phys. Rev. A} \textbf{105}, 063325 (2022).

\bibitem{bereta2021} Bereta, S. J., Caracanhas, M. A. \& Fetter, A. L. Superfluid vortex dynamics on a spherical film. \textit{Phys. Rev. A} \textbf{103}, 053306 (2021).

\bibitem{padavic2020} Padavi{\'c}, K., Sun, K., Lannert, C. \& Vishveshwara, S. Vortex-antivortex physics in shell-shaped Bose-Einstein condensates. \textit{Phys. Rev. A} \textbf{102}, 043305 (2020).

\bibitem{vitelli2004} Vitelli, V. \& Turner, A. M. Anomalous Coupling Between Topological Defects and Curvature. \textit{Phys. Rev. Lett.} \textbf{93}, 215301 (2004). 

\bibitem{blakie2008} Blakie, P. B., Bradley, A. S., Davis, M. J., Ballagh, R. J. \& Gardiner, C. W. Dynamics and statistical mechanics of ultra-cold Bose gases using c-field techniques. \textit{Adv. Phys.} \textbf{57}, 363-455 (2008).

\bibitem{proukakis2008} Proukakis, N. P. \& Jackson, B. Finite-temperature models of Bose–Einstein condensation. \textit{J. Phys. B: At. Mol. Opt. Phys.} \textbf{41}, 203002 (2008).

\bibitem{ho2015} Ho, T.-L. \& Huang, B. Spinor Condensates on a Cylindrical Surface in Synthetic Gauge Fields. \textit{Phys. Rev. Lett.} \textbf{115}, 155304 (2015).

\bibitem{helm2015} Helm, J. L., Cornish, S. L. \& Gardiner, S. A. Sagnac Interferometry Using Bright Matter-Wave Solitons. \textit{Phys. Rev. Lett.} \textbf{114}, 134101 (2015).

\bibitem{zhang2018} Zhang, J. \& Ho, T.-L. Potential scattering on a spherical surface. \textit{J. Phys. B: At. Mol. Opt. Phys.} \textbf{51}, 115301 (2018).

\bibitem{ouvry2019} Ouvry, S. \& Polychronakos, A. P. Anyons on the sphere: Analytic states and spectrum. \textit{Nucl. Phys. B} \textbf{949}, 114797 (2019). 

\bibitem{shi2017} Shi, Z.-Y. \& Zhai, H. Emergent gauge field for a chiral bound state on curved surface. \textit{J. Phys. B: At. Mol. Opt. Phys.} \textbf{50}, 184006 (2017).

\bibitem{bloch2012} Bloch, I., Dalibard, J. \& Nascimbène, S. Quantum simulations with ultracold quantum gases. \textit{Nat. Phys.} \textbf{8}, 267–276 (2012).

\bibitem{georgescu2014} Georgescu, I. M., Ashhab, S. \& Nori, F. Quantum simulation. \textit{Rev. Mod. Phys.} \textbf{86}, 153 (2014).

\bibitem{hu2019} Hu, J. et al. Quantum simulation of Unruh radiation. \textit{Nat. Phys.} \textbf{15}, 785–789 (2019).

\bibitem{munozdenova2019} Mu\~{n}oz de Nova, J. R. et al. Observation of thermal Hawking radiation and its temperature in an analogue black hole. \textit{Nature} \textbf{569}, 688–691 (2019).

\bibitem{viermann2022} Viermann, C. et al. Quantum field simulator for dynamics in curved spacetime. \textit{Nature} \textbf{611}, 260–264 (2022).

\bibitem{visser2001} Barcel\'{o}, C., Liberati, S. \& Visser, M. Analogue gravity from Bose-Einstein condensates. \textit{Class. Quantum Grav.} \textbf{18}, 1137 (2001).

\bibitem{banik2022} Banik, S. et al. Accurate Determination of Hubble Attenuation and Amplification in Expanding and Contracting Cold-Atom Universes. \textit{Phys. Rev. Lett.} \textbf{128}, 090401 (2022).

\bibitem{gomezllorente2019} Gomez Llorente, J. M. \& Plata, J. Expanding ring-shaped Bose-Einstein condensates as analogs of cosmological models: Analytical characterization of the inflationary dynamics. \textit{Phys. Rev. A} \textbf{100}, 043613 (2019).

\bibitem{bhardwaj2021} Bhardwaj, A., Vaido, D. \& Sheehy, D. E. Inflationary dynamics and particle production in a toroidal Bose-Einstein condensate. \textit{Phys. Rev. A} \textbf{103}, 023322 (2021).

\bibitem{eckel2021} Eckel, S. \& Jacobson, T. Phonon redshift and Hubble friction in an expanding BEC. \textit{SciPost Phys.} \textbf{10}, 064, (2021).

\bibitem{cooper2008}  Cooper, N. R. Rapidly rotating atomic gases. \textit{Adv. Phys.} \textbf{57}, 539-616 (2008).

\bibitem{fetter2009} Fetter, A. L. Rotating trapped Bose-Einstein condensates. \textit{Rev. Mod. Phys.} \textbf{81}, 647 (2009).

\bibitem{ho2001} Ho, T.-L. Bose-Einstein Condensates with Large Number of Vortices. \textit{Phys. Rev. Lett.} \textbf{87}, 060403 (2001).

\bibitem{mukherjee2022} Mukherjee, B. et al. Crystallization of bosonic quantum Hall states in a rotating quantum gas. \textit{Nature} \textbf{601}, 58-62 (2022).

\bibitem{schweikhard2004} Schweikhard, V., Coddington, I., Engels, P., Mogendorff, V. P. \& Cornell, E. A. Rapidly Rotating Bose-Einstein Condensates in and near the Lowest Landau Level. \textit{Phys. Rev. Lett.} \textbf{92}, 040404 (2004).

\bibitem{streubel2016} Streubel, R. et al. Magnetism in curved geometries. \textit{J. Phys. D: Appl. Phys.} \textbf{49}, 363001, (2016).

\bibitem{gentile2022} Gentile, P., et al. Electronic materials with nanoscale curved geometries. \textit{Nat. Electron.} \textbf{5}, 551–563 (2022). 

\bibitem{guinea2010} Guinea, F., Katsnelson, M. \& Geim, A. Energy gaps and a zero-field quantum Hall effect in graphene by strain engineering. \textit{Nat. Phys.} \textbf{6}, 30–33 (2010).


\end{thebibliography}
\end{document}